\documentstyle[12pt,aaspp4,flushrt]{article}
\newcommand{\beq}{\begin{equation}}              % end equation
\newcommand{\eeq}{\end{equation}}             % end equation
\newcommand{\beqa}{\begin{eqnarray}}              % end equation array
\newcommand{\eeqa}{\end{eqnarray}}             % end equation array
\def\msun{{\rm\,M_\odot}}
\def\hmsun{{h^{-1}\rm\,M_\odot}}

\def\mpc{{\rm\,Mpc}}
\def\hmpc{{h^{-1}\rm\,Mpc}}

\def\kms{{\rm\,km\,s^{-1}}}

\def\etal{et al.\ }

\let\lsim=\lesssim
\def\rta{R_{ta}}
\begin{document}
\title{Dynamical Mass Estimates of Large-Scale Filaments \\
In Redshift Surveys}
\author{Daniel J. Eisenstein\altaffilmark{1}
\altaffiltext{1}{Also at: Physics Department, Harvard University}
and Abraham Loeb}
\affil{Astronomy Department, Harvard University,\\ 
60 Garden St., Cambridge MA 02138}
\author{Edwin L. Turner}
\affil{Department of Astrophysics, Princeton University,\\
Peyton Hall, Princeton, NJ 08544}

\begin{abstract}
We propose a new method to measure the mass of large-scale filaments in
galaxy redshift surveys.  The method is based on the fact that the mass
per unit length of isothermal filaments depends only on their
transverse velocity dispersion.  Filaments that lie perpendicular to
the line of sight may therefore have their mass per unit length
measured from their thickness in redshift space.  We present
preliminary tests of the method and find that it predicts the mass per
unit length of filaments in an N-body simulation to an accuracy of
$\sim\!35\%$.  Applying the method to a select region of the
Perseus-Pisces supercluster yields a mass-to-light ratio of $M/L_B \approx
460h$ in solar units to within a factor of two.  The
method measures the mass-to-light ratio on length scales of up to
$\sim\!50\hmpc$ and could thereby yield new information on the behavior
of the dark matter on mass scales well beyond that of clusters of
galaxies.

\bigskip
\noindent
\end{abstract}
%%% \keywords{}
\bigskip\bigskip
\begin{center}
Submitted to {\it The Astrophysical Journal.}
\end{center}

\clearpage
\section{Introduction}

It has long been known that galaxies are not spread evenly throughout
the universe but instead are organized into larger structures
stretching out to scales $\sim\!100\mpc$ (\markcite{Ein80}Einasto \etal
1980; \markcite{Pee93}Peebles 1993).  Large redshift surveys
(\markcite{Str95}see Strauss \& Willick 1995 for a review) mapped this
structure in three dimensions and showed that in addition to the
conspicuous clusters of galaxies, there are also extended
one-dimensional filaments (\markcite{Hay86}Haynes \& Giovanelli 1986)
and two-dimensional sheets (\markcite{deL86}de Lapparent \etal 1986;
\markcite{Gel89}Geller \& Huchra 1989; \markcite{She96}Shectman \etal
1996).  The observed geometries can be explained by gravitational
instability theories of structure formation, both through analytical
approximations (\markcite{Zel70}Zel'dovich 1970; \markcite{Bon95}Bond
\etal 1995; \markcite{Eis95}Eisenstein \& Loeb 1995), and numerical
simulations (e.g.\ \markcite{Ber91}Bertschinger \& Gelb 1991;
\markcite{Par94}Park \etal 1994; \markcite{Cen94}Cen \& Ostriker
1994).

The filaments and sheets which are observed in galaxy surveys represent
the most massive nonlinear structures in the local universe. While
these structures are painted by the light emitted from galaxies, their
actual mass distribution is unknown.  Determining the mass is
particularly difficult because these systems are still evolving along
one or two axes.
Less massive systems such as galaxies and galaxy clusters have
generally virialized by now, and their dynamics unambiguously implies
substantial amounts of dark matter (\markcite{Rub83}Rubin 1983;
\markcite{Tri87}Trimble 1987; \markcite{Dav95}David \etal 1995).
While the dynamical estimates of the mass-to-light ratio of virialized
systems argue for an open universe (e.g. \markcite{Bah95}Bahcall \etal
1995), it is unknown whether more mass, sufficient to close the
universe, remains undetected outside virialized objects.  Methods to
measure the mass on larger scales include peculiar velocity studies
(see \markcite{Str95}Strauss \& Willick 1995 and references within),
analyses of superclusters (\markcite{Pos88}Postman \etal 1988;
\markcite{Ray91}Raychaudhury \etal 1991; \markcite{Baf93}Baffa \etal
1993), application of the cosmic virial theorem (\markcite{Dav83}Davis
\& Peebles 1983), and the inferences based on cosmic microwave
background anisotropies (\markcite{Jun95}Jungman \etal 1995).  In
addition to the implications for the value of the cosmic density
$\Omega$, such measurements provide insight into the clustering
properties of the dark matter and the degree of biasing in galaxy
formation.

In this paper, we present a novel method for measuring the mass of large
scale filaments that is based purely on the information available in
redshift surveys.  The method relies on the observation that while
spherical and planar geometries require both a characteristic velocity and
a characteristic length to estimate mass, a cylindrical system requires
only a velocity dispersion to estimate its mass per unit length.  On
dimensional grounds, the mass per unit length times Newton's constant
must be proportional to the transverse velocity dispersion squared of the
filament.  For filaments oriented across the sky, i.e.  perpendicular to
the line of sight, the velocity dispersion is measured as the thickness of
the structure in redshift space.  Since the length of such a filament is
readily apparent from its angular extent, the method allows the
determination of the mass-to-light ratio on scales beyond that of clusters,
despite the fact that the objects of interest are neither fully virialized
nor in the linear perturbative regime.

In \S2, we present an exact solution to the Jeans equation for the case
of an isothermal, axisymmetric, steady-state filament.  This derivation
extends the well-known hydrodynamic solutions for isothermal gases
(\markcite{Sto63}Stod\'{o}\l kiewicz 1963; \markcite{Ost64}Ostriker
1964) to collisionless systems.  With this solution at hand, we present
tests of the method in \S3, focusing primarily on filaments selected in
real space from a N-body simulation.  While not definitive, the results
are encouraging, suggesting that accuracy $\sim\!30\%$ in mass is
attainable.  In \S4 we conclude with a discussion of the ingredients of
more elaborate scheme to calibrate the method and demonstrate its
robustness.
We also apply the method as it stands to the Perseus-Pisces supercluster
and estimate a $B$-band mass-to-light ratio of $460h$ in solar units.

\section{Analytic Results}
Let us first derive an exact analytical solution to the Jeans equations
for the case of an axisymmetric, isothermal, steady-state filament that
is translationally invariant along its symmetry axis.  Solutions for
isothermal gaseous filaments in hydrostatic equilibrium
(\markcite{Sto63}Stol\'{o}\l kiewicz 1963; \markcite{Ost64}Ostriker
1964) lead the way.  We begin with the Jeans equations in cylindrical
coordinates ($R$, $\theta$, $z$) and assume axial symmetry and no bulk
velocity in the radial (transverse) direction. The radial Jeans
equation is,
\beq \label{eq2a}
{\partial(\nu \overline{v^2_R})\over\partial R} + \nu {\overline{v^2_R}
- \overline{v^2_\theta}\over R} = -\nu{\partial \Phi\over\partial R};
\eeq
where $\nu$ is the number density of particles, $\overline{v^2_R}$ 
and $\overline{v^2_\theta}$ are the ensemble averages of the squares
of the radial velocities and tangential velocities, respectively,
and $\Phi$ is the gravitational potential.  All of these quantities
are functions of $R$ only.  Gravity is determined from
the Poisson equation,
\beq \label{eq2b}
{1\over R}{\partial\over\partial R}\left(R{\partial\Phi\over\partial R}\right)
= 4\pi G\rho,
\eeq
where $G$ is Newton's constant and $\rho\propto\nu$ is the mass density.

We now assume that $\overline{v^2_R}$ and $\overline{v^2_\theta}$ are
related by a constant $\beta =
1-\overline{v^2_\theta}/\overline{v^2_R}$.  As in the case of analysis
of spherical systems (\markcite{Bin87}Binney \& Tremaine 1987),
$\beta=1$ indicates purely radial orbits, $\beta=0$ indicates isotropic
orbits, and $\beta=-\infty$ indicates purely tangential orbits.  The
introduction of $\beta$ reduces equation (\ref{eq2a}) to
\beq \label{eq2c}
{1\over\nu}{\partial(\nu\overline{v^2_R})\over\partial R} +
\beta {\overline{v^2_R}\over R} = -{\partial\Phi\over\partial R}.
\eeq
We now assume that the filament is isothermal, so that
$\overline{v^2_R}(R) = \sigma^2$.  This yields
\beq \label{eq2d}
\sigma^2\left({\partial\log\rho\over\partial\log R}+\beta\right) = 
-R{\partial\Phi\over\partial R}.
\eeq
Note that this equation is scale-free in radius.

We next introduce the mass per unit length enclosed within a radius $R$
\beq \label{eq2e}
\mu(R) = 2\pi\,\int^R_0 \tilde R~\rho(\tilde R)\,d\tilde R.
\eeq
Inserting (\ref{eq2b}) yields
\beq \label{eq2f}
\mu = {1\over2G}R{\partial\Phi\over\partial R},
\eeq
which in turn may be used in equation (\ref{eq2d}) to yield
\beq \label{eq2g}
{\sigma^2\over2G}\left({R\over\rho}\rho'+\beta\right) = -\mu,
\eeq
where primes indicate differentiation with respect to $R$.
From equation (\ref{eq2e}), we find $\mu'=2\pi R\rho$, which we
use to eliminate $\rho$ in favor of $\mu$.  The resulting equation is
\beq \label{eq2h}
R\mu'' + (\beta-1)\mu' + {2G\over\sigma^2}\mu\mu' = 0
\eeq
Since $R\mu'' = d(R\mu'-\mu)/dR$, we may integrate (\ref{eq2h}) to get
\beq \label{eq2i}
R \mu' + (\beta-2)\mu + {G \mu^2\over\sigma^2} = 0;
\eeq
since $\mu(0)=0$, the constant of integration is zero.
We may then integrate again and solve for $\mu$ to find
\beq \label{eq2j}
\mu (R) = 
(2-\beta){\sigma^2\over G}{R^{2-\beta}\over R^{2-\beta}+R_0^{2-\beta}},
\eeq
where $R_0$ is an arbitrary scale factor.  Therefore, the mass at radii
much larger than $R_0$ is finite and approaches $(2-\beta)\sigma^2/G$.

We may differentiate $\mu(R)$ to find the density profile,
\beq \label{eq2k}
\rho = {(2-\beta)^2 \sigma^2\over2\pi G R_0^2}
{x^{-\beta}\over(x^{2-\beta}+1)^2},
\eeq
where $x=R/R_0$.  Hence, $\rho\propto x^{-\beta}$ at small radii and
$\rho\propto x^{\beta-4}$ at large radii.  The small radii behavior is
unphysical for $\beta<0$ (i.e.  predominantly tangential orbits).

Finally, we consider what velocity dispersion is measured along a line
of sight perpendicular to the axis of symmetry.  
Assuming that the filament is axisymmetrically sampled,
we find that the measured 1-d velocity dispersion orthogonal to the 
symmetry axis is
\beq \label{eq2m}
\sigma^2_{\perp} = {\overline{v^2_R}+\overline{v^2_\theta}\over2}.
\eeq
Inserting our assumptions concerning the ratio of the velocity dispersions
and isothermality, this becomes
$\sigma^2_{\perp} = \sigma^2(1-\beta/2)$.
Hence, we find that the total mass per unit length of the filament
(defined for $R\gg R_0$) is 
\beq \label{eq2n}
\mu = {2\sigma_\perp^2\over G} = 7.4\times10^{13}\msun\mpc^{-1}\,
\left(\sigma_\perp\over 400\kms\right)^2.
\eeq
The dependence on $\beta$ has canceled out.

\section{Numerical Tests on Real-space Filaments}
Based on the discussion in \S 2, we propose to use the observed velocity
dispersion as a means of calibrating the mass per unit length of a filament
of galaxies in a galaxy redshift surveys.  Here one would study filaments
that are aligned perpendicular to the line of sight; the velocity
dispersion then manifests itself as the thickness of the filament in
redshift space.  By measuring the mass per unit length of such structures,
one can find their mass-to-light ratios and thereby probe the properties of
dark matter on large scales.  However, the analytic results of the last
section were derived under a particular set of idealized assumptions.  The
filaments of galaxies in a redshift survey do not satisfy all these
assumptions, and therefore the validity of equation (\ref{eq2n}) needs to
be checked against one-dimensional structures in numerical simulations.

The most drastic violation of the assumptions underlying \S2 is due to
omnipresent substructure, often in the form of fragmentation along the
filament (\markcite{Cha53}Chandrasekhar \& Fermi 1953;
\markcite{Sto63}Stod\'{o}\l kiewicz 1963; \markcite{Lar85}Larson
1985).  What one takes as a filament actually more resembles a chain of
differently-sized beads.  Substructure tends to cause an overestimation
of the mass per unit length, essentially because vacant areas along the
filament are credited with having mass when in fact they are empty.
For example, if a ``filament'' were actually a string of $N$
widely-spaced isothermal spheres of velocity dispersion $\sigma$ and
radius $R$, then the true mass would be $2\sigma^2NR/G$.  But the
filamentary mass estimate is $2\sigma^2L/G$ where $L$ is the length of
the filament, an overestimate by a factor of $L/NR$, or twice the
filling fraction of the isothermal spheres.

Another key difference is that the filaments are not isolated but instead
are subject to continuing infall and perturbations from neighboring mass
concentrations. The transverse crossing time across the filament is much
shorter than the Hubble time, and so the filament core may virialize.
However, the infalling material violates the assumption of steady-state
radial equilibrium; moreover, due to the associated redshift distortion
(\markcite{Sar77}Sargent \& Turner 1977; \markcite{Kai87}Kaiser 1987),
it makes the filament look thinner (i.e.\ have a
lower $\sigma$) than it actually is.  Self-similar infall solutions for
filamentary geometries have been found in both collisionless and
collisional systems (\markcite{Fil84}Fillmore \& Goldreich 1984;
\markcite{Inu92}Inutsuka \& Miyama 1992, \markcite{Geh96}Gehman \etal
1996), and so one could imagine deriving the equivalent of equation
(\ref{eq2n}) for these collapse solutions.  However, because the mass
inside a particular radius diverges as the radius increases in these
infall models, it is unclear how to define the total mass per unit
length of a redshift space filament.  Moreover, if the background
cosmology is not scale-free, then the self-similar solution loses its
justification.  All solutions, infalling or isolated, will produce
$\mu\propto\sigma^2/G$ by dimensional analysis, and so we see no
compelling reason to disfavor the coefficient of 2 found in the
isolated case relative to other approximations. Instead we calibrate
this coefficient using N-body simulations.

Finally, real filaments are not infinitely long, exactly straight, or
perfectly isothermal and axisymmetric.  The effects of finite length or
curvature may be characterized by a length scale, either the length or
radius of curvature, which are then compared to the characteristic width of
the filament.  In either case, the fact that this length scale is larger
than the length scale we expect to be associated with fragmentation
suggests that these effects will be smaller than the deviations caused by
fragmentation.  The isothermal assumption has worked well in spherical
systems, but remains to be tested in this case.  Deviations from
axisymmetry will cause variations in the inferred mass per unit length as a
function of viewing angle; we will estimate the magnitude of the
variations later in this section.

We see two methods for testing the applicability of equation (\ref{eq2n}).
First, we may select filamentary structures in real space from N-body
simulations.  This procedure utilizes information that is not available
in redshift surveys, but it does allow us to test whether this dynamical
mass estimation formula holds for systems that stretch the idealizations
under which it was derived.  In particular, we may investigate the
role of substructure within the filament.  Second, we may select the
filaments in redshift space from mock surveys culled from simulations.
This allows one to examine the effects of contamination from foreground
and background galaxies, to experiment with selection effects, and to
calibrate the method in a robust way.  We focus on the first of these
methods in this paper, although we will devote some discussion to the
second.

For our testing, we use an open CDM particle-mesh (PM) N-body simulation
provided by C. Park and J.R.\ Gott (\markcite{Par94}Park \etal 1994).
The simulation has $240^3=13.8\times10^6$ particles and a $480^3$ mesh;
the background cosmology is $\Omega=0.4$, $\Lambda=0$, and
$H_0=50\,{\rm km\,s^{-1}\,Mpc^{-1}}$.  The simulation volume is $576
h^{-3}\mpc^3$, yielding a particle mass of $1.5\times10^{12}\hmsun$,
where $h=H_0/(100\,{\rm km\,s^{-1}\,Mpc^{-1}}) = 0.5$.  We use only the
$z=0$ output.  Typical filaments have masses around
$3\times10^{15}\hmsun$ and lengths of order $50\hmpc$.  They are
generally a few, but rarely more than 10, mesh cells thick; the mass
per unit length is such that there are on average 30 particles per
lengthwise mesh spacing.

\subsection{Real-space Selected Filaments}
For our tests on real-space selected filaments, we select filaments
by eye from cross-sectional slices.
We look for candidates which appear to be roughly linear arrangements
of particles.  By looking at orthogonal cross-sections, we
verify that the object is indeed a filament rather than a chance
superposition.  We then choose two endpoints of a line segment to
describe the center of the filament; the filament is defined as a
cylindrical volume of a particular radius around this line segment.

The choice of the filament's boundary radius is not unique.
While one might expect the steady-state assumption in the idealized
derivation to apply only to the densest regions, where the particles have
executed several radial crossings of the filament, this is not the region
which will be picked out in a redshift survey.  Redshift-space distortions
will cause objects infalling onto the filament to be confused with those in
the collapsed region.  For example, a particle at turnaround has zero
radial velocity relative to the center of the filament and therefore has
the same redshift.  To pick a radius characteristic of this infall region,
we select the radius at which the average density within that cylinder is
5.7 times the background density.  
This is the value for the density at
turnaround of a collapsing homogeneous filament in an $\Omega=0.4$
universe (for reference, the value would be 3.5 for $\Omega=1$ and 8.9 for
$\Omega=0.2$).  
As we show later,  our density definition indeed selects
the turnaround radius of actual filaments in an N-body simulation.
We denote this radius by $\rta$.

Small filaments may be underresolved by the PM code; we therefore require
that within the radius $\rta$ the filament contains at least 500 particles
and has an average linear density exceeding 12.5 particles per length-wise
mesh spacing.  This leaves us with 23 filaments; the largest ones are 6
times more massive than the minimum mass requirement.

We next consider ``observing" these particles from a direction orthogonal
to the filament axis.  Because we are interested in the velocity dispersion
of the particles rather than in their bulk motions, we do not want
variations in bulk velocity along the filament to be included in its
redshift ``thickness".  In particular, a filament that is orthogonal to the
viewer's line of sight in real space may be slightly tilted or warped in
redshift space and we do not want such variations to enter the velocity
dispersion.  Therefore, we break the filament lengthwise into 20 equal
pieces and remove the mean velocity in each piece from the velocity of the
particles in the section before calculating the velocity dispersion.

Taking the axis of the filament to be the $z$ axis, we consider 
observing the filament from directions in the $x$-$y$ plane.  Different
viewing angles yield different estimates for the line-of-sight velocity
dispersion; however, these varying answers are in fact only different
combinations of $\sigma_{xx}$, $\sigma_{yy}$, and $\sigma_{xy}$,
where for example 
\beq\label{eq3a}
\sigma^2_{xy} = {1\over N}\sum_{particles}v'_x v'_y;
\eeq
$v'_i$ is the $i$-th component of the velocity of a particle after
the mean velocity of its section of the filament has been subtracted,
and $N$ is the total number of particles.
Forming the matrix
\beq\label{eq3b}
\Sigma = \left(\begin{array}{cc}\sigma^2_{xx}&\sigma^2_{xy}\\
\sigma^2_{xy}&\sigma^2_{yy}\end{array}\right)
\eeq
and considering an observer at infinite distance from direction ${\bf \hat{n}}$
in the $x$-$y$ plane, the velocity dispersion in that direction is
simply $\sqrt{{\bf\hat{n}}^T\Sigma{\bf\hat{n}}}$.  Therefore, by diagonalizing
$\Sigma$, we find the largest and smallest possible measurements of
$\sigma_\perp^2$ that can result from differing viewing angles.
The figures show these two extremal estimates for the
mass per unit length; the ratio between them is usually $\lsim\!2$.

In Figure 1, we show the comparison for the 23 filaments between the
true mass per unit length within $\rta$, $\mu_{true}(\rta)$, and the two
extreme estimates $\mu_{est}(\rta)$ based on the measurement of  
$\sigma^2_\perp$.  Figure 2 shows $\mu_{true}(\rta)$ versus the ratio
$\tilde{\mu}=\mu_{est}(\rta)/\mu_{true}(\rta)$.  Here we see
that the estimate is in almost all cases within a factor of 2 from
the true value, and with this sample there is no obvious correlation 
between $\mu_{true}$
and $\tilde{\mu}$.  Taking all viewing angles as equally likely, this
ensemble of 23 filaments yields a distribution of $\tilde{\mu}$ with a
mean of 1.17 and a $1$-$\sigma$ error of 0.39.  In other words, the
estimate of $\mu$ from equation (\ref{eq2n}) is biased 17\% high; with
this removed, one finds a mass estimate with 33\% accuracy.

Because the filament is laid out across the sky, the substructure and
clumpiness along its length are observable.  We would expect that
applying our method to filaments with more substructure would produce
larger estimates of the mass per unit length, essentially because one is
crediting the lower-density regions with the velocity dispersion of the
higher-density regions.  We consider several different statistics to
measure the degree of substructure.  First, we break the filament into
20 lengthwise pieces, count the number of particles in each piece, and
take the ratio of the standard deviations of these 20 numbers to their
mean as one statistic.  Next, we bin the particles into 128 lengthwise
bins and perform the cosine transform (\markcite{Pre92}Press \etal
1992).  We then add up the power in the 10, 15, or 20 lowest modes
(normalizing away the dependence on the number of bins and the number
of particles) and use these as measures of substructure.

In Figure 3, we plot the power in the 10 lowest cosine modes, $S_{10}$,
versus the ratio $\tilde{\mu}$ of the estimated to true mass per unit
length.  There is a fair correlation ($r=0.53$) in the expected
direction.  The other substructure statistics produce very similar results.
In applying this program to real survey data, one might plan to reject
filaments with large measures of substructure.  If we remove the 6
filaments with $S_{10}$ larger than 0.9, then the remaining 17
filaments produce a distribution of $\tilde{\mu}$ with mean 1.07 and
error 0.31 (29\%).

Next, we examine the dependence of the velocities on radius.  In Figure
4, we show the azimuthally-averaged profiles for the mean radial
velocity $\overline{v_R}(R)$ and radial velocity dispersion
$\sigma_R(R)$ for several particular filaments.  We also show the
average value of $\sigma_\perp$ that can be measured from the particles
within the given radius.  The profiles $\overline{v_R}(R)$ show that
$\rta$ is indeed a reasonable choice for the turn-around radius.  The
profiles of $\sigma_R(R)$ show deviations from isothermality at large
radii; because $\sigma_\perp$ is a cumulative statistic, these
deviations do not affect the measured velocity dispersion much.  For
the full sample of 23 filaments, the ratio of $\sigma_R(\rta)$ to the
value of $\sigma_R$ for all particles within $\rta/2$ has a mean of
0.80 with error 0.247; that is, the velocity dispersion at $\rta$ is
roughly 80\% of the value in the central regions.  Finally, we find
that the average value of the velocity anisotropy parameter $\beta$ is
zero (isotropic) at all radii but with significant scatter
(1-$\sigma\sim0.3$).

In summary, the method of predicting the mass per unit length of a filament
from its observed velocity dispersion does reasonably well in real-space
tests of an N-body simulation. Because the procedures of this section did
not confront the confusion caused by redshift distortions and
foreground/background galaxies or the missteps possible in picking
one-dimensional structures from a discrete set of points, we do not consider
these tests definitive. Nevertheless, it is encouraging that the method
performs to an accuracy of $\lsim\!40\%$ despite significant substructure,
departures from axisymmetry and isothermality, and some inclusion of
infall.

\subsection{A First Step into Redshift Space}
We next apply the method to a simulated redshift survey.  Using the
same PM simulation as above, we select a mock survey by giving each
particle an equal chance to be a galaxy, assuming a Schechter
luminosity function, and applying an apparent magnitude cutoff.  The
survey geometry is chosen to be a slice $1.5^\circ$ thick and
$90^\circ$ wide, and the depth is similar to that of the Las Campanas
Redshift Survey (LCRS) (\markcite{She96}Shectman \etal 1996;
\markcite{Lin96}Lin \etal 1996; \markcite{Lan96}Landy \etal 1996).  We
use the adjacent slices to verify that our filaments are not cuts
through sheets.  The real-space plot of the particles in shown in
Figure 5; the plot of the mock redshift survey drawn from the slice is
shown in Figure 6.

Within the slice, two large transverse filaments are apparent.
We flag the ``galaxies'' which appear as part of the filament.  The
galaxies selected this way are highlighted in Figure 6.  Within a set of
galaxies, we fit a straight line in redshift space and measure the
residual velocity spread around the line to find the velocity
dispersion $\sigma_\perp$.  Because the filaments have rather little
extent in redshift, we neglect variations in the selection function
across the set and give each galaxy equal weight in the velocity
dispersion.  We also neglect the small angle between the observed line
of sight and the direction perpendicular to the axis of the filament.
More refined analyses could include these effects.  Our measurement of
$\sigma_\perp$ then gives the mass per unit length, which we multiply
by the length (angular length times distance) to get the total mass.

In order to compare the mass estimate to the true answer, we find it most
convenient to convert our result to an estimate of $\Omega$.  This is possible
because we know the luminosity function that was assumed to apply to
all particles in the simulation.  $\Omega$ is then the mass-to-light
ratio of the filament times the ratio of the luminosity density (as
derived from the luminosity function) to the critical density.  This
measured $\Omega$ may then be compared to the true value in the
simulation, $\Omega=0.4$.

Filament A has 296 galaxies at a distance of 22,000 km/s.  We measure a
velocity dispersion $\sigma_\perp=400\kms$ and a length of $110\hmpc$.
This yields a mass of $7.9\times10^{15}\hmsun$ and $\Omega=0.65$.
Filament B has 305 galaxies at a distance of 14,000 km/s.  We measure
$\sigma_\perp=450\kms$ and a length of $40\hmpc$, yielding a mass of
$3.6\times10^{15}\hmsun$ and $\Omega=0.58$.

These two cases therefore yield overestimates of $\Omega$ by about
50\%.  The agreement between the estimated mass and the true mass is
surprisingly good, bearing in mind the somewhat ambiguous choice of the
member galaxies of the filament.  In fact, by restricting ourselves to
a $1.5^\circ$ slice, we may have clipped out some galaxies that would
have been included in the filament in a less 2-dimensional survey; such
galaxies lie just above or below the slice.  Because the ``fingers of
God'' are preferentially in the slice rather than above or below, these
additional galaxies would most likely not increase the velocity
dispersion but would increase the ``light'' associated with the
filament, thereby reducing the estimated value of $\Omega$.  This
problem, combined with the difficulties in rejecting sheets, suggests
that thin slices like the LCRS are less appropriate for this method
than surveys with wider sky coverage such as the Sloan Digital Sky
Survey (SDSS) (\markcite{Gun93}Gunn \& Knapp 1993).
The two cases presented here are merely intended to be illustrative; we
discuss what is required for a full calibration and testing of the method
in the next section.

\section{Discussion and Conclusion}
Motivated by the observation that the mass per unit length of an isothermal
filament depends only on its velocity dispersion and not on its scale
radius, we have proposed a method to measure the dynamical mass of
large-scale filaments in galaxy redshift surveys.  A filament aligned
across the plane of the sky would have its velocity dispersion easily
observable as the thickness of the filament in redshift space.  The degree
of substructure along the filament is also observable and might be used to
warn against filaments that drastically violate the validity regime of
our method.  In a wide-angle redshift survey, such as CfA/SSRS
(\markcite{Vog94}Vogeley \etal 1994; \markcite{daC94}da Costa \etal
1994), 2-degree Field (\markcite{Tay95}Taylor 1995),
or SDSS (\markcite{Gun93}Gunn \& Knapp 1993), it should be possible to distinguish
between one-dimensional filaments and two-dimensional sheets.  The
method could then be used to estimate the masses and mass-to-light
ratios of these structures, which are up to 10 times more massive than
rich clusters, and thus yield information on the behavior of the dark
matter on these scales.

The bulk of the tests we present are performed on filaments selected from
cross-sectional real-space slices of an $\Omega=0.4$ simulation.  We pick
the radius of the filament to be 
the turn-around radius, here found as the radius at which the
enclosed density is 5.7 times the background density.  We find that the
velocity dispersion of those particles, as measured from a direction
perpendicular to the axis of the filament, is a good predictor of the
mass contained within that radius.  The ensemble of 23 filaments so
treated yields an estimate which is biased $\sim\!20\%$ high with a
scatter $\sim\!33\%$ (1-$\sigma$).

However, a full test of the robustness of the method must be done in
redshift space and not in real space.  
To illustrate the situation in redshift space, we have analyzed two case
studies (cf.\ Fig.\ 5).  However, we leave to a future paper the larger task
of designing an automated Monte Carlo scheme to calibrate the method and
prove its robustness.  Such a program would begin from a cosmological
simulation and extract a mock redshift survey.  One would then apply a
quantitative algorithm to this survey in order to extract filamentary
structures.  With these filaments in hand, one would compute the velocity
dispersion and convert it to a mass-to-light ratio or $\Omega$.  By
considering many systems, one may calibrate the method and find the error
in its estimates, possibly as a function of some substructure criteria.
One could then repeat this using different cosmological simulations in
order to show that the calibration is indeed robust against variations in
the cosmological model.

Because the steady-state assumption used in the derivation of equation
(\ref{eq2n}) does not hold in cosmological situations, the coefficient
of 2 in equation (\ref{eq2n}) is subject to further calibration.  Since
the amount of infall depends upon the background cosmology, it is
likely that a given calibration will not be unbiased in all
cosmologies.  In particular, the higher degree of present-day infall in
$\Omega=1$ models might cause systematic underestimates of the mass
relative to a low $\Omega$ calibration.  It will be important to
quantify this effect so as to assess the method's ability to
differentiate between cosmological models.  Cylindrical self-similar
solutions (\markcite{Fil84}Fillmore \& Goldreich 1984) or
secondary-infall solutions (\markcite{Got75}Gott 1975;
\markcite{Gun77}Gunn 1977; \markcite{Hof85}Hoffman \& Shaham 1985;
\markcite{Ryd87}Ryden \& Gunn 1987) might provide analytic testbeds for
understanding the relation between the virialized and infall regions
and for probing the dependence of the results on the underlying
cosmology.

The density-morphology relation of galaxies
(\markcite{Dre80}\markcite{Dre84}Dressler 1980, 1984;
\markcite{Pos84}Postman \& Geller 1984) may produce a significant
systematic effect on our mass estimate, when combined with the
selection criteria of an actual survey.  Because elliptical galaxies
prefer high-density regions and spirals conversely, a survey that
favors one type over the other will weight the regions of high and low
velocity dispersion differently, thereby producing differing estimates
of $\mu$ (see, e.g.  differences between estimates of the bias factor
in optical and IRAS surveys (\markcite{Pea94}Peacock \& Dodds 1994)).
In extreme cases, the skewed selection function may affect the
performance of the filament-finding algorithm.  Similarly, a survey
such as the LCRS which undersamples close pairs would affect the
calculation of $\mu$.  Including a parameterized form of the
density-morphology relation in the extraction of the mock surveys might
allow one to treat the systematic differences between redshift
surveys.

In this paper, we assumed that the galaxy distribution followed the
mass distribution of the simulation and used simulation particles as
galaxy tracers.  Biased galaxy formation could significantly affect the
validity of this treatment.  Velocity bias (\markcite{Car89}Carlberg \&
Carlberg 1989; \markcite{Cou92}Couchman \& Carlberg 1992;
\markcite{Car94}Carlberg 1994; \markcite{Sum95}Summers \etal 1995) may
cause the velocity of galaxies to be diminished relative to the dark
matter, especially in denser regions.  Estimates of this effect vary
but are often $\sim\!20\%$ for clusters; however, a comparison between
optical and X--ray observations of clusters seems to indicate that the
bias is less than a 10\% effect (\markcite{Lub93}Lubin \& Bahcall
1993).  In addition there is the possibility that some aspects of
biasing are associated with the filamentary structure itself, e.g.\ if
fragmentation of filaments affects galaxy formation.  This would not be
the case in a hierarchical structure formation model, in which the
galaxy-scale perturbations should collapse well before the large-scale
structure forms, leaving the galaxies to fall onto the filaments in a
manner similar to the dark matter.  Numerical work on velocity bias
have focused on clusters, but could be extended to the filamentary
case.

The Perseus-Pisces supercluster is the obvious nearby candidate to
which to apply the method, as its central ridge is a prominent linear
structure stretching across the sky in redshift space
(\markcite{Hay86}Haynes \& Giovanelli 1986).  To estimate the
mass-to-light ratio, we use the data from the \ion{H}{1}\ redshift
survey of Giovanelli, Haynes, and collaborators (\markcite{Weg93}Wegner
\etal 1993 and references therein). We focus on a restricted portion of
the central ridge with (RA, $\delta$) corners of ($0^{\rm h}30^{\rm
m}$, $27^\circ$), ($2^{\rm h}15^{\rm m}$, $36.5^\circ$), ($2^{\rm
h}15^{\rm m}$, $40.5^\circ$), and ($0^{\rm h}30^{\rm m}$, $31^\circ$)
in order to avoid the heavily extincted region near the Perseus
cluster.  We impose a heliocentric velocity cut of
$4000\kms<v<5800\kms$; the remaining galaxies have a velocity
dispersion of $430\kms$ and an average distance of $51\hmpc$.  From
this we infer a mass per unit length
$\mu=8.5\times10^{13}\msun\mpc^{-1}$, a length of $21\hmpc$, and a mass
$M=1.8\times10^{15}\hmsun$.  Assuming that all of the selected galaxies
are at a distance of $51\hmpc$, imposing a uniform extinction
correction of 0.2 mag (\markcite{Gio86}Giovanelli \etal 1986), and
assuming based on integrating the luminosity function
(\markcite{Mar95}Marzke 1995) that the survey includes 56\% of the
light, we estimate the total $B$-band luminosity to be
$3.9\times10^{12}h^{-2}$ in solar units.  Hence, we find $M/L_B\approx
460h$ in solar units; however, the effects described earlier in this
section render this estimate uncertain within a factor of two.  Further
refinements of the method and its application to additional filaments,
as will be available with surveys such as the SDSS (Gunn \& Knapp
1993), should substantially reduce these uncertainties.

\bigskip
We thank Changbom Park and J.\ Richard Gott for kindly supplying the
results of their numerical simulation and Riccardo Giovanelli and
Martha Haynes for supplying their redshift compilation in electronic
form.  We thank Neta Bahcall, Karl Fisher, Bob Kirshner, Tsafrir
Kolatt, Huan Lin, Jordi Miralda-Escude, and Eve Ostriker for helpful
discussions.  D.J.E. was supported by a National Science Foundation
Graduate Reseach Fellowship.  A.L.  was supported in part by the NASA
ATP grant NAG5-3085.

%%% Include these lines if one only wants figure captions.
%%% Comment out all the EPSF include lines.
%%% \clearpage
%%% \begin{center}
%%% {\bf Figure Captions}
%%% \end{center}

\clearpage\centerline{\epsfxsize=\textwidth\epsffile{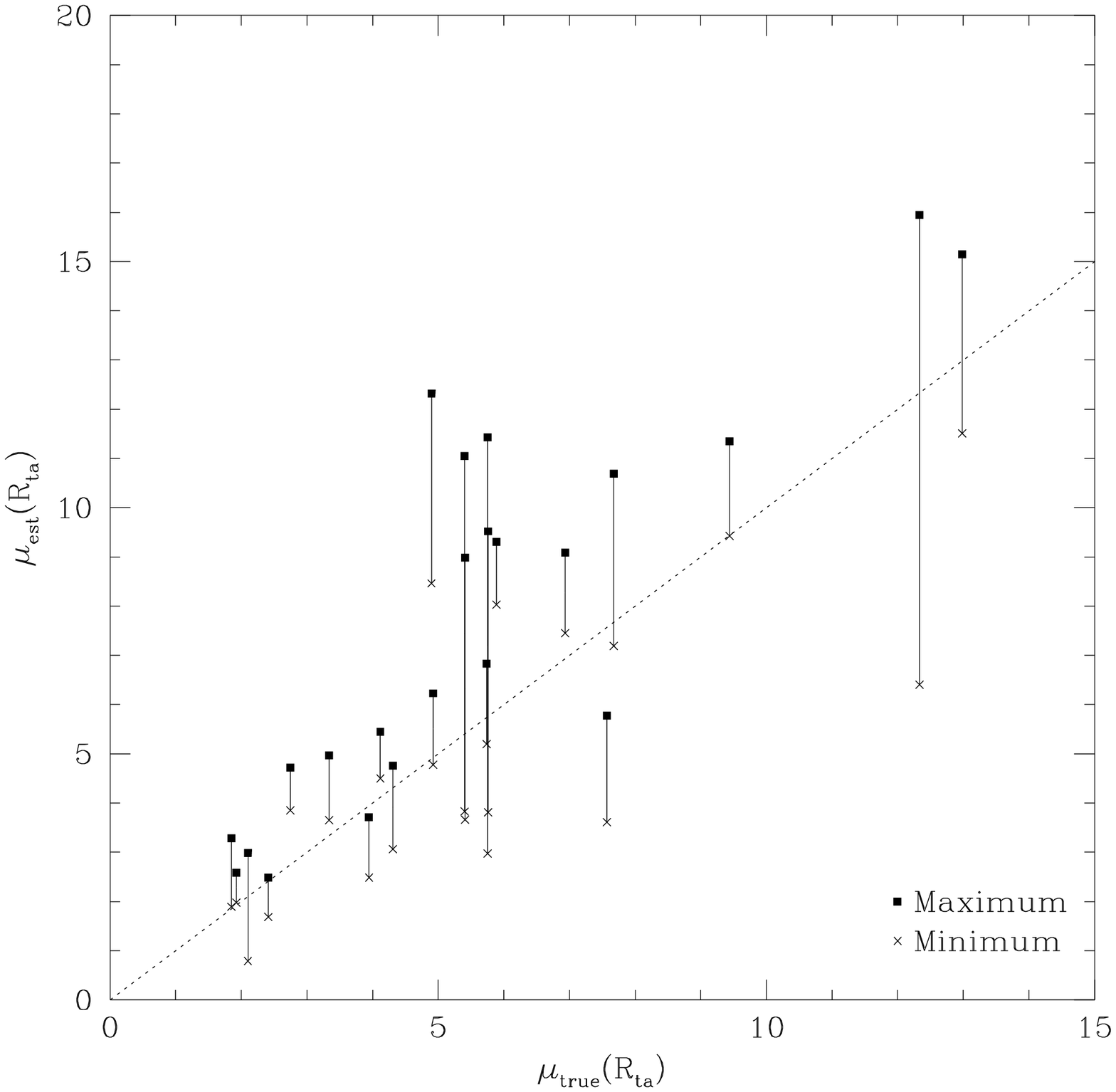}}
\bigskip\noindent{\bf Figure 1:}
The mass per unit length as estimated from the velocity dispersion of
the particles within radius $\rta$ plotted against the actual mass per
unit length within radius $\rta$.  The plot shows the full range of
possible estimates if one considers all viewing angles in the plane
perpendicular to the axis of the filament.

\clearpage\centerline{\epsfxsize=\textwidth\epsffile{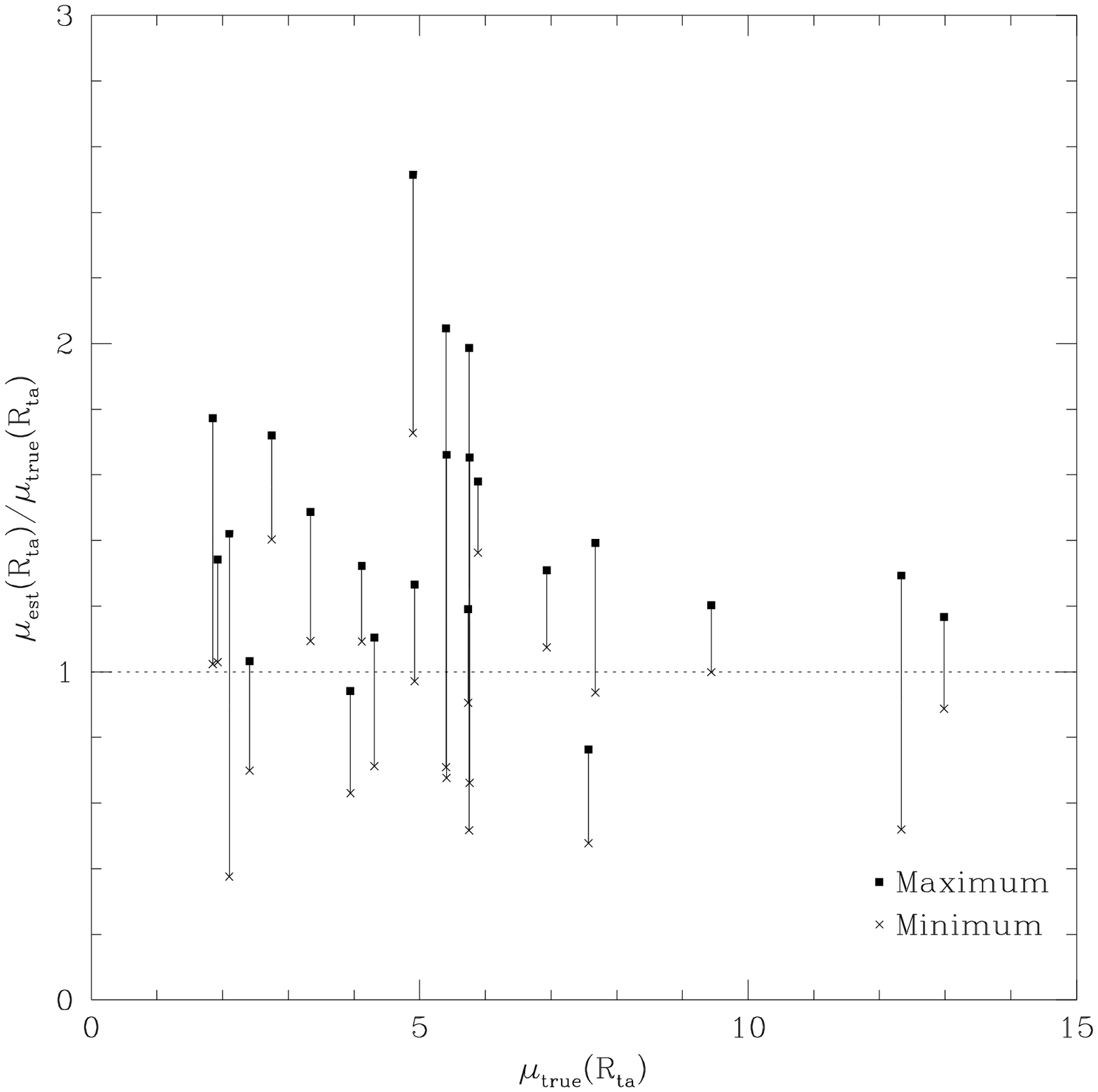}}
\bigskip\noindent{\bf Figure 2:}
As Figure 1, but shown is the ratio of the estimated mass
per unit length $\mu$ to the true value versus the true $\mu$.

\clearpage\centerline{\epsfxsize=\textwidth\epsffile{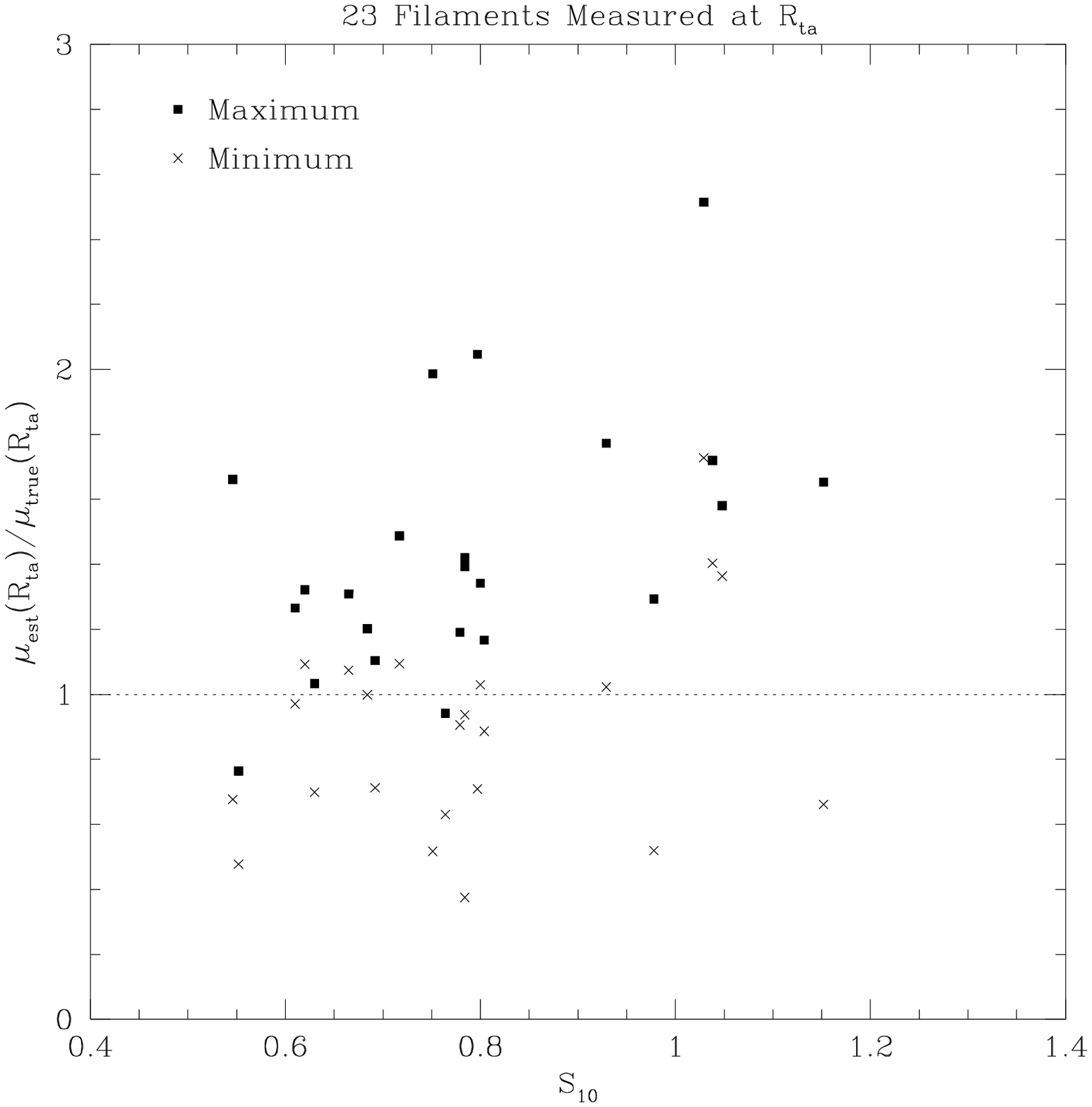}}
\bigskip\noindent{\bf Figure 3:}
The ratio of the estimated mass per unit length to the true value
versus a measure of substructure within the filament.  The statistic
$S_{10}$ is the sum of the power in the lowest 10 modes of the cosine 
transform; larger values indicate more substructure.

\clearpage\centerline{\epsfxsize=\textwidth\epsffile{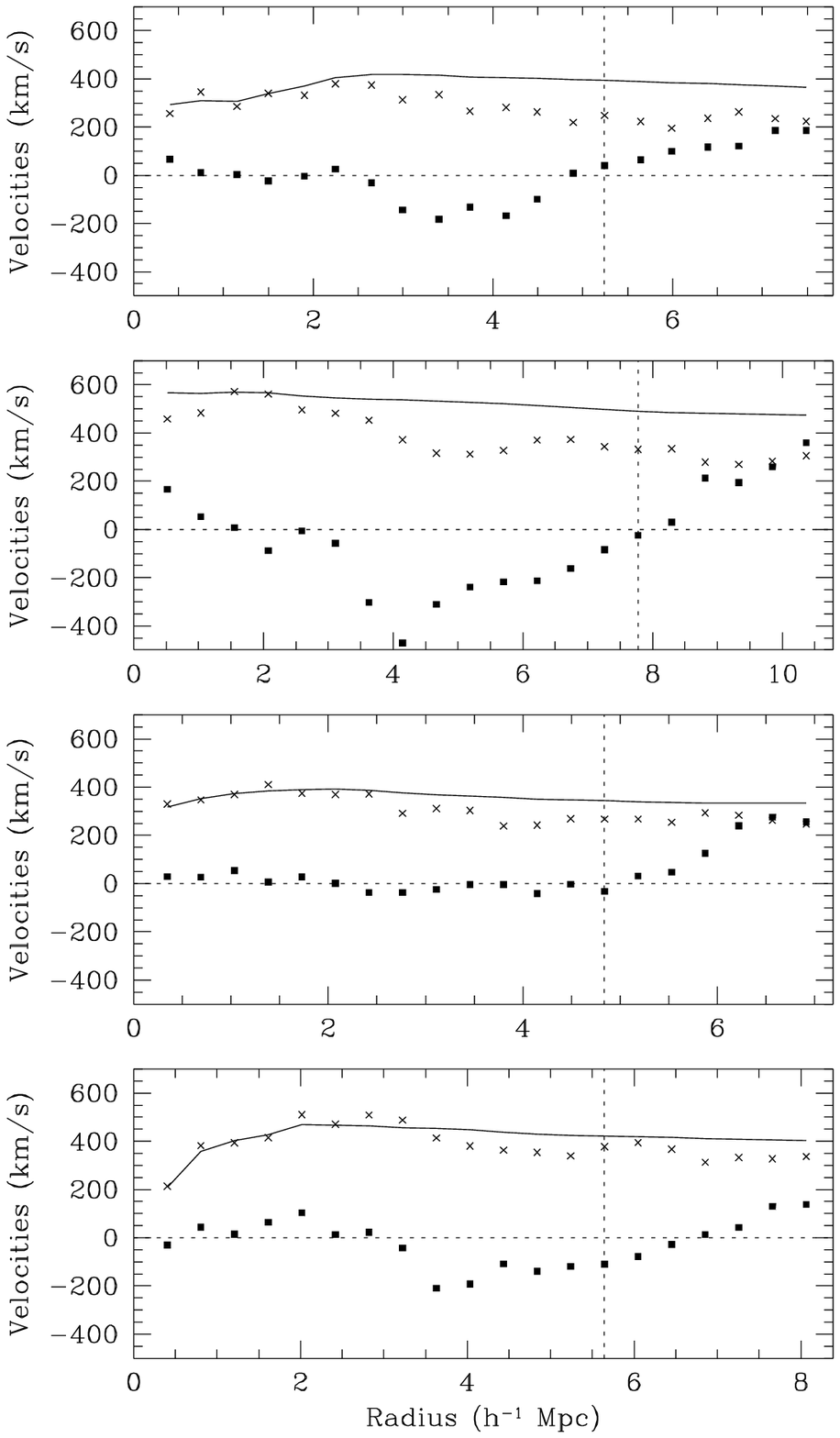}}
\bigskip\noindent{\bf Figure 4:}
The velocity profiles of 4 filaments from the sample.  For each radial bin,
we show the the mean radial velocity ({\it squares}) and the radial
velocity dispersion ({\it crosses}).  Also shown is the average transverse
velocity dispersion of all particles within the stated radius ({\it solid
line}).  The vertical dashed lines mark $\rta$ in each case. The third and
fourth panels show the regions marked in Figure 5 as filament A and B,
respectively.

\clearpage\centerline{\epsfxsize=\textwidth\epsffile{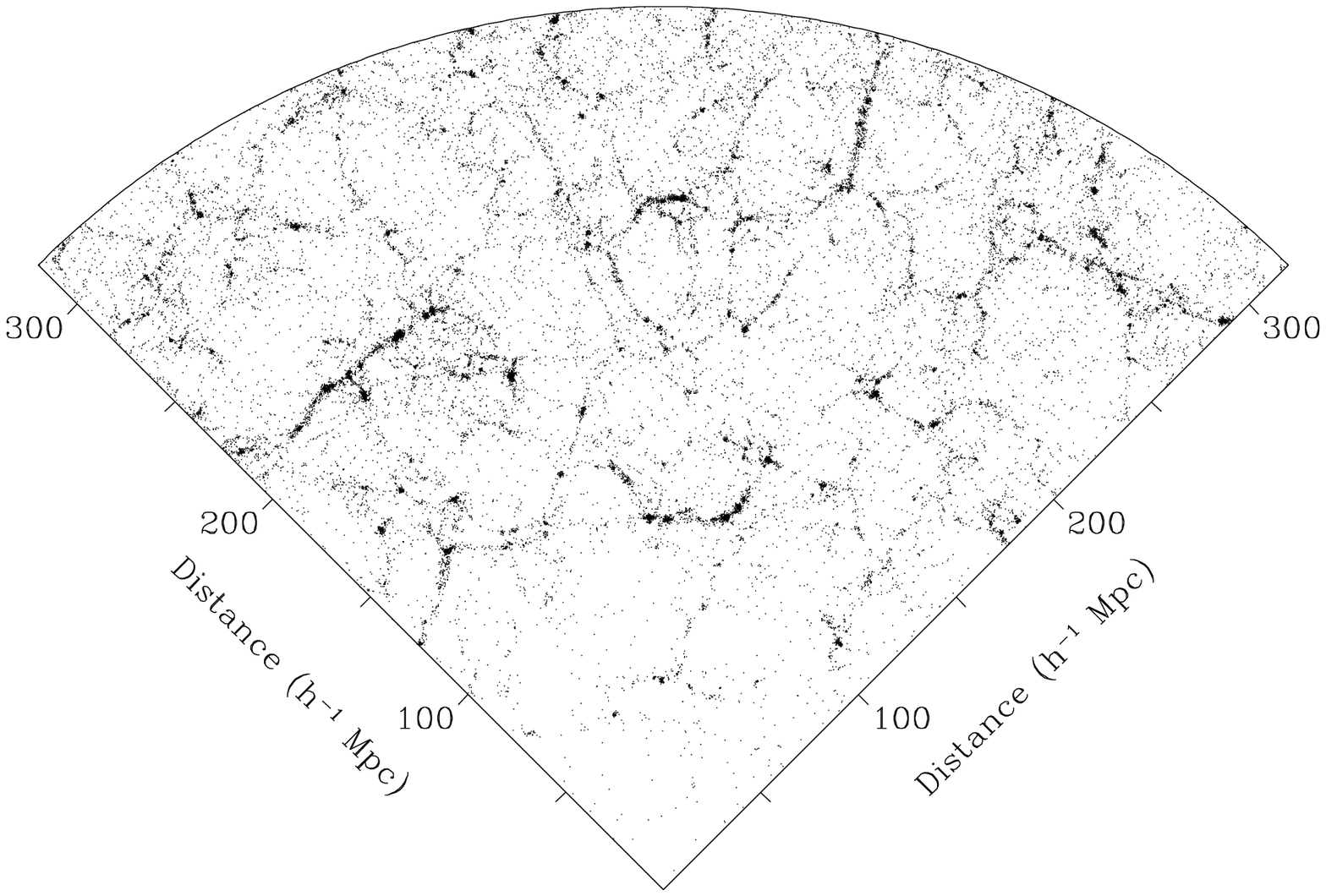}}
\bigskip\noindent{\bf Figure 5:}
A slice of the simulation displaying the particle positions.  The slice
is $1.5^\circ$ thick.

\clearpage\centerline{\epsfxsize=\textwidth\epsffile{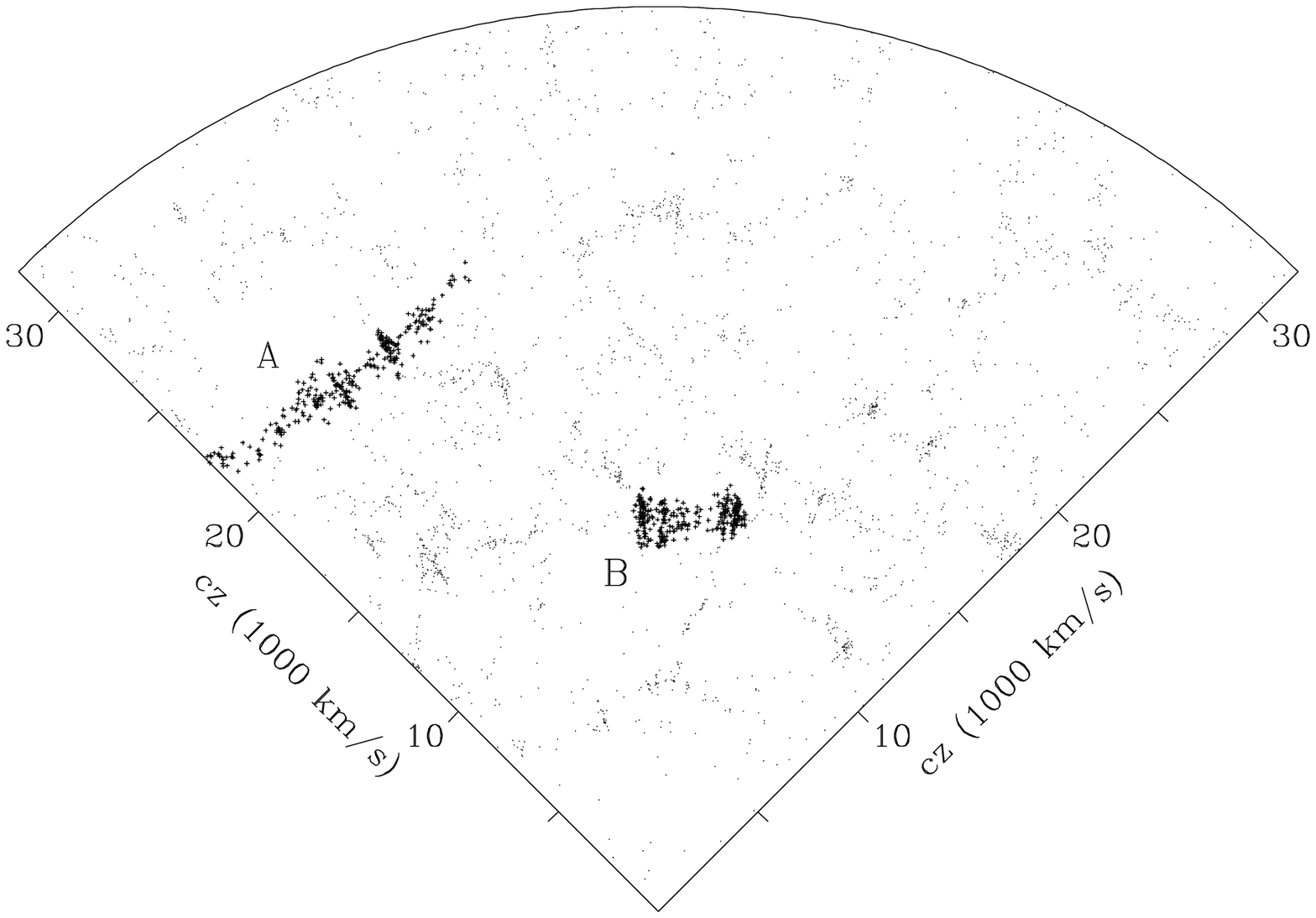}}
\bigskip\noindent{\bf Figure 6:}
A mock redshift survey drawn from the slice displayed in Figure 5.  The
two filaments analyzed in the text are marked with the letters A and B.

\end{document}